\documentclass[preprint, aps, draft, pre]{revtex4}
\usepackage[dvips]{graphicx}

\begin{document}

\title{Block Copolymer at Nano-Patterned Surfaces}

\author{Xingkun Man$^{1}$, David Andelman$^{1}$ and Henri Orland$^{2}$}
\affiliation{$^{1}$Raymond and Beverly Sackler School of Physics and Astronomy,
Tel Aviv University, Ramat Aviv 69978, Tel Aviv, Israel\\
$^{2}$Institut de Physique Th\'eorique, CEA-Saclay,
F-91191 Gif-sur-Yvette Cedex, France}

\date{Ver 9, after refereeing, June 29, 2010}

\begin{abstract}

We present numerical calculations of lamellar
phases of block copolymers at patterned surfaces.
We model symmetric di-block copolymer films forming lamellar
phases and the effect of geometrical and chemical surface patterning on the alignment and orientation
of lamellar phases. The calculations are done within self-consistent field theory (SCFT), where the
semi-implicit relaxation scheme is used to solve the diffusion equation.
Two specific set-ups, motivated by recent experiments, are investigated. In the first, the film is
placed on top of a surface imprinted with long chemical stripes. The stripes interact
more favorably with one of the two blocks and induce a perpendicular orientation in a large range
of system parameters.
However, the system is found to be sensitive to its initial conditions, and sometimes gets trapped
into a metastable mixed state composed of domains  in parallel and perpendicular orientations.
In a second set-up, we study the film structure and orientation
when it is pressed against a hard grooved mold. The mold surface prefers one of the two components
and this set-up is found to be superior  for inducing a perfect perpendicular lamellar orientation
for a wide range of system parameters.

\end{abstract}

\maketitle

\section{Introduction} \label {sec.1}

Block copolymers (BCP) have been studied extensively in the last few decades
due to their special self-assembly properties
giving rise to interesting mesophases in the sub-micrometer to nanometer range,
as well as to their numerous applications, where desired
properties can be tailored by specific chain architecture \cite{HamleyBook,r1,r2,r3,r4,r5,r6,r7}.

Bulk properties of BCP are rather well understood and, in recent years, much effort was devoted
to understand thin films of BCP. One potential application is the use of thin films of
di-block copolymers
as templates and scaffolds for the fabrication of arrays of nanoscale domains, with high control over
their long-range ordering, and with the hope that this technique can be useful in future micro-
and nano-electronic applications. Recent experiments include
using chemically \cite{r8,r9,r10,r11,r12,r13,r14,r15,r15a,r15b} and physically
\cite{r16,r17,r18,r19} patterned surfaces, which have preferential local wetting properties for one of the
two polymer blocks. The orientation and alignment of lamellar and hexagonal phases of BCP were investigated,
and, in particular, their transition between parallel (`lying down') and perpendicular (`standing up') orientations. Another useful method
is the use of electric fields to orient anisotropic phases of BCP, such as lamellar and hexagonal,
in a direction perpendicular to the solid surface \cite{r20,r21,r22,r23,r24,r25,r26,r27,r28}.

In this paper, we present self-consistent field theory (SCFT) calculations  inspired by recent
experiments on patterned surfaces \cite{r9,r11,r12,r15a,r15b,r29,r29a,r29b}. Our main aim is to analyze what are the thermodynamical conditions
that facilitate the perpendicular orientation of BCP lamellae with respect to the underlying solid surface,
and how the lamellar ordering can be optimized. Two specific solid patterns and templates are modeled.
The first is a  planar solid surface that has a periodic arrangement of long and parallel stripes
preferring one of the two blocks, but otherwise is neutral to the two blocks in its inter-stripe regions.
We show that this experimentally realized surface pattern \cite{r9,r11,r12,r15a,r15b} enhances the perpendicular
lamellar orientation. The second surface pattern is motivated by recent NanoImprint lithography (NIL)
experiments~\cite{r15b,r29,r29a,r29b}. This is a high-throughput low-cost process which has the potential of
reducing the need for costly surface preparation. Here, a hard grooved mold is pressed onto a thin BCP
film at temperatures above the film glass-transition and induces perpendicularly oriented lamellae.
Within our model we show that, indeed, the grooved surface does enhance the perpendicular orientation of lamellae.

The SCFT model that we use in the BCP calculations has several known limitations. It is a coarse-grained
model and, as such, can only describe spatial variations that are equal or larger than the monomer size
(the Kuhn length). Our calculations provide the thermodynamical equilibrium, or local minima of the film
free-energy in presence of geometrical constraints. Therefore, important structural details
induced by hydrodynamic flow and film rheology as occurring during sample preparation
are not described by the model.

In the present work we limit ourselves to three-dimensional systems
that are translationally invariant along one spatial direction. This is applicable when the BCP film is put in
contact with surfaces having long unidirectional stripes or grooves. Extensions of the present work to more
complex three-dimensional systems with two-dimensional surface patterns will be addressed separately in a follow-up publication.

The outline of our paper is as follows. In the next section two system set-ups are introduced;
a chemically striped surface and a grooved mold and their effect on orienting lamellar BCP films is presented.
In Sec. III we describe our SCFT model and how its equations are solved numerically. In Sec. IV our results are
presented for the two types of experimental set-ups. Finally, in the last section we discuss the model
predictions and their connection with experimental findings.

\section{The BCP Film design}

We consider a melt of A-B  di-block
copolymer (BCP) chains composed of $n$ chains, each having a length $N=N_A+N_B$ in terms of the Kuhn length $a$ that is assumed,
for simplicity, to be the same for the A and B monomers. Hence, the A-monomer molar fraction $f=N_A/N$ is equal to its volume fraction.
In addition, hereafter we concentrate on symmetric di-BCP, $N_A=N_B$ having $f=0.5$. The symmetric BCP yields thermodynamically stable
lamellar phases of periodicity $\ell_0$, as the temperature is lowered below the order-disorder temperature (ODT) \cite{ft1}.
At shallow temperature quenches, simple scaling arguments~\cite{r37} used  in the weak segregation
limit show that the lamellar period $\ell_0$ is proportional to $R_g$, the chain radius of gyration,
$\ell_0\sim R_g=\sqrt{Na^2/6}\sim N^{1/2}$. For deep temperature quenches well below the ODT, the
strong segregation theory~\cite{r39} yields more stretched chains as $\ell_0\simeq N^{2/3} \gg R_g\sim N^{1/2}$.

The BCP film has total volume  $\Omega$ and lateral area $\cal{A}$, so that its   thickness is
$L=\Omega/\cal{A}$. In some experimental set-ups  the BCP film is bounded by two planar solid surfaces, and its thickness $L$ is a
constant.  In other set-ups \cite{r9,r11,r12,r15a,r15b} the film is spin coated on a solid surface with a free polymer/air interface on
its top, so that the thickness can vary spatially. In yet another set-up used in NanoImprint lithography (NIL) experiments
\cite{r15b,r29,r29a,r29b}, a grooved mold is pressed against the film and the film penetrates into the mold. As the film profiles
inside the mold varies considerably in height, $L$ is only the film average thickness.

We will consider only surface features along one spatial direction (chosen to be the $x$-direction),
and assume that the system is translationally invariant along the second surface direction (the $y$-direction). Hence,
the film volume $\Omega$ (per unit length)
has units of length square, while the surface area $\cal{A}$ has units of length. The third spatial direction, the $z$
one, is taken to be perpendicular to the surfaces.  This allows us to carry out  the numerical calculations only in the
($x$,$z$) two-dimensional plane, and represents a considerable simplification from the numerical point of view.

The situation where a thin BCP film is placed in contact with a flat and uniform surface (or is sandwiched between two flat surfaces)
was modeled by several authors~\cite{r30,r31,r32,r33,r34,r34a,r35,r36,r37,r38}.
Two main features are apparent when the film behavior is compared to that of bulk BCP. The first effect is the film confinement.
When $L$ differs from the natural periodicity  $\ell_0$, the chains need to be stretched or compressed as the film is
incompressible and space filling. The film free-energy shown in Fig.~\ref{fig1} is a function of the thickness $L$,
and is obtained within our SCFT scheme (see below), and
agrees well with previous results \cite{r30,r34a}.

\begin{figure}[htp]
  \begin{center}
    {\resizebox{3.5in}{!}{\includegraphics[angle=0,scale=0.7,draft=false]{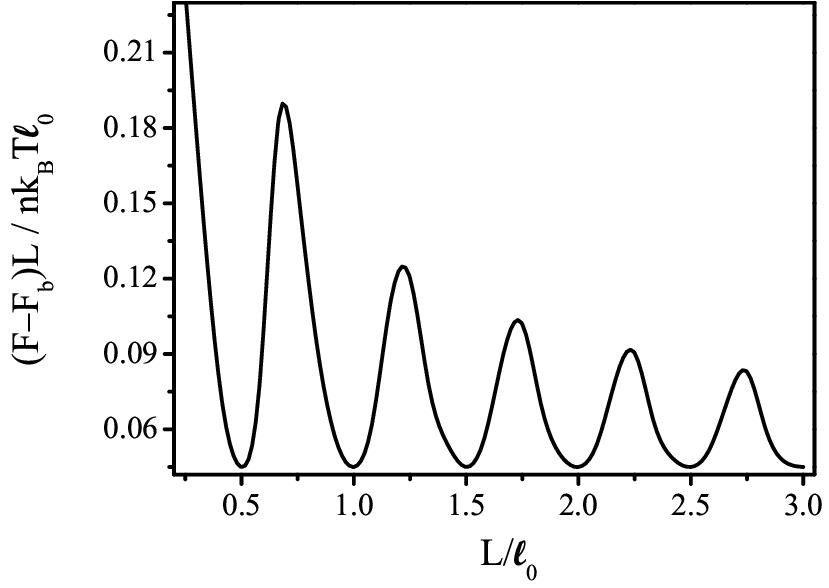}}}
    \caption{\textsf{The difference of the dimensionless film free-energy from its bulk value, $(F-F_b)L/(n k_B T \ell_0)$,
    as function of the film
    rescaled thickness, $L/\ell_0$, where $\ell_0$ is the lamellar periodicity, $k_B T$ is the thermal energy
    and $n$ the total number of chains. The lamellae in the film are parallel to the two flat bounding surfaces
    (the ${\rm L}_{||}$ state). The surface preference is $\Delta u=1$ for the bottom surface and $\Delta u=0$ for the top, and $N\chi=20$.
    \label{fig1}}}
  \end{center}
\end{figure}

The main effect of the confinement between the two bounding surfaces is the existence of
free-energy minima at integer or half-integer values of  $L/\ell_0=\frac{1}{2},1,\frac{3}{2},...$ corresponding
to film thicknesses where we can fill an integer or half-integer numbers of A-B
parallel layers in between the two surfaces. The overall trend for the film free-energy is to converge toward the bulk value $F_b$ as: $F-F_b\sim 1/L$.

\begin{figure}[htp]
  \begin{center}
    {\resizebox{3.5in}{!}{\includegraphics[angle=0,scale=0.7,draft=false]{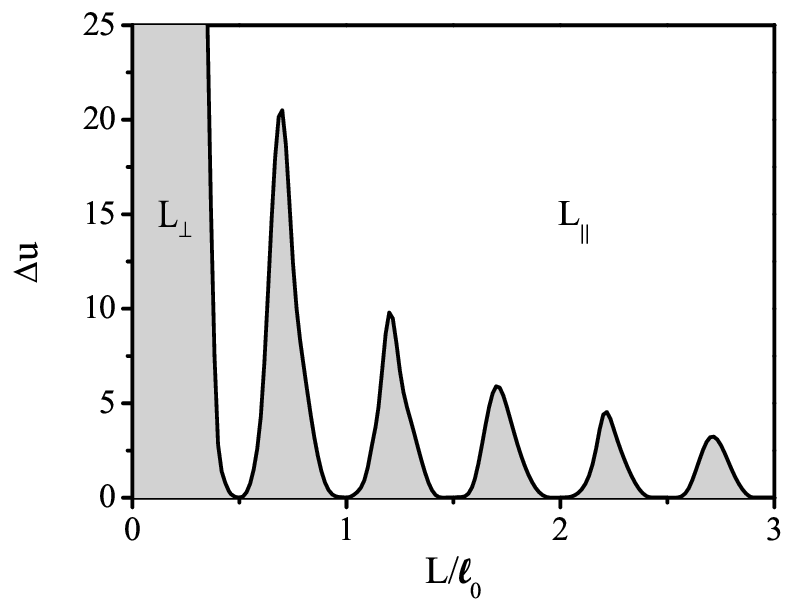}}}
    \caption{\textsf{Phase diagram for intermediately segregated ($N\chi =20$) symmetric di-block
lamellar phase in terms of the film thickness $L/\ell_0$ vs the surface field
 preference between the two blocks, $\Delta u=u_A-u_B$. The grey area indicates the perpendicular
 lamellar region (${\rm L}_\perp$) and the white region above it is the parallel state (${\rm L}_{||}$).
 The film is confined in the $z$-direction
between two parallel and flat surfaces.
The bottom surface uniformly attracts one of blocks ($\Delta u>0$), while
the top one mimics the free (and neutral) surface, $\Delta u=0$. \label{fig2}}}
  \end{center}
\end{figure}

The second feature is the possibility to induce a parallel to perpendicular
transition of the lamellae by changing the strength of the surface preference, $\Delta u$.
This can be seen in Fig.~\ref{fig2} where the parallel to perpendicular phase diagram is plotted
in the $\Delta u$ -- $L$ plane, within our SCFT scheme. When a strong surface preference towards
one of the two blocks is included, the lamellae tend to orient in a parallel direction, while for
 neutral (indifferent) surfaces or weak preferences, the perpendicular orientation is preferred as the lamellae can
assume their natural periodicity $\ell_0$ in this orientation for any thickness $L$. Note also
that the transition occurs at $\Delta u=0$ for integers or half-integers values:
$L/\ell_0=\frac{1}{2},1,\frac{3}{2},...$ as was argued above. These results agree well with those reported in Refs.~\cite{r30,r34a,r37}.

In the remaining of the paper we will address in detail the question of how
it is possible to better control the relative stability of
parallel and perpendicular phases of lamellar BCP films. And, in particular, how can the stability of the
perpendicular phase be increased for a larger range of film thicknesses and surface characteristics.

\section{Theoretical Framework} \label {sec. 3}

Since the system is translationally invariant in the $y$-direction,
we treat it as an effective two-dimensional system. The free
energy for such a di-block copolymer (BCP) film confined between the
two surfaces is
\begin{eqnarray}\label{f1}
\frac{a^2}{k_{B}T}{F} &=&\int {\rm d}^{2} r
\left[\chi\phi_{A}(r)\phi_{B}(r)-\omega_{A}(r)\phi_{A}(r)-\omega_{B}(r)\phi_{B}(r)\right]\nonumber\\
 & &-{na^2}\ln Q_{C}-\int {\rm d}^2{r} \left[u_{A}(r)\phi_{A}(r)+u_{B}(r)\phi_{B}(r)\right]
 \nonumber\\
 & & +\int {\rm d}^2 r\, \eta(r)[\phi_A(r)+\phi_B(r)-1]
\end{eqnarray}
where each of the $n$  BCP chains is composed of $N=N_A+N_B$ Kuhn
segments of length $a$, and the
Flory-Huggins parameter is $\chi$.
The dimensionless volume
fractions of the two components are defined as $\phi_{A}(r)=\phi_A(x,z)$ and
$\phi_B(r)=\phi_B(x,z)$, respectively, whereas $\omega_{j}(r)$, $j={\rm A, B}$,
are the auxiliary fields coupled with $\phi_{j}(r)$, and $Q_{C}$ is
the single-chain partition function in the presence of the $\omega_A$ and $\omega_B$ fields
[see Eqs.~(\ref{f3a})-(\ref{f3}) below for more details].
The third term represents a surface energy preference, where $u_A$ and $u_B$
are the short-range interaction parameters of the surface with the A and B monomers,
respectively. Formally, $u_A(r)$ and $u_B(r)$ are surface fields and get non-zero values
only on the surface(s).

Finally, the last term includes a
Lagrange multiplier $\eta(r)$ introduced to ensure the
incompressibility condition of the BCP melt:
\begin{equation}
\phi_{A}(r)+\phi_{B}(r)= 1 \quad {\rm for~~ all} \quad r\in \Omega
\label{f2a}
\end{equation}
By inserting this condition, Eq.~(\ref{f2a}), in the surface free energy of Eq.~(\ref{f1}),
the integrand becomes
$u_A\phi_A+u_B\phi_B=(u_A-u_B)\phi_A +u_B$. Hence,
$\Delta u(r)\equiv u_A(r)-u_B(r)$ is the only needed surface preference
field that will be employed throughout the paper.

Using the saddle-point approximation, we obtain a set of
self-consistent equations
\begin{eqnarray} \label{f2}
\frac{\delta F}{\delta\phi_{A} }=0~\Rightarrow  ~~~\omega_{A}(r)&=&\chi\phi_{B}(r)-u_{A}(r)+\eta(r)\nonumber\\
\frac{\delta F}{\delta\phi_{B} }=0~\Rightarrow  ~~~\omega_{B}(r)&=&\chi\phi_{A}(r)-u_{B}(r)+\eta(r)\nonumber\\
\frac{\delta F}{\delta\omega_{A} }=0~\Rightarrow  ~~~\phi_{A}(r)&=
&\frac{n a^2}{\Omega Q_{C}}\int^{N_{A}}_{0} ds\,
q_{A}(r,s)q^{\dag}_{A}(r,N_{A}-s)\nonumber\\
\frac{\delta F}{\delta\omega_{B} }=0~\Rightarrow
~~~\phi_{B}(r)&=&\frac{n a^2}{\Omega Q_{C}}\int^{N_{B}}_{0} ds\, q_{B}(r,s)q^{\dag}_{B}(r,N_{B}-s)
\end{eqnarray}
where the incompressibility condition, Eq.~(\ref{f2a}), is obeyed,
and the single-chain free energy $Q_c$ is:
\begin{equation}\label{f3a}
Q_{C}=\frac{1}{\Omega}\int {\rm d}^2r \,q^{\dag}_{A}(r,N_{A})
\end{equation}
The two types of propagators $q_{j}\left(r,s\right)$ and
$q^{\dag}_{j}\left(r,s\right)$ (with $j={\rm A, B}$) are solutions
of the modified diffusion equation
\begin {equation} \label{f3}
\frac{\partial q_{j}(r,s)}{\partial s}
=\frac{a^{2}}{6}\nabla^{2}q_{j}\left(r,s\right)-\omega_{j}(r)q_{j}(r,s)
\end {equation}
with the initial condition $q_{A}(r,s{=}0)\,{=}\,q_B(r,s{=}0)\,{=}\,1$,
$q^{\dag}_{A}(r,s{=}0)\,{=}\,q_{B}(r,N_{B})$ and
$q^{\dag}_{B}(r,s{=}0)\,{=}\,q_{A}(r,N_{A})$, where $s$ is a conveniently defined curvilinear
coordinate along the chain contour. This diffusion equation
is solved using reflecting boundary conditions at the two
confining surfaces  ($z=0$ and $z=L$): ${\rm d} q/{\rm d} r|_{z=0}\,{=}\,0$
and ${\rm d} q/{\rm d} r|_{z=L}\,{=}\,0$, while  periodic  boundary conditions are used in the perpendicular direction.

Hereafter, we rescale all lengths by the natural periodicity of
the BCP, $\ell_{0}\simeq 4.05R_{g}$, \cite{ft2} where
$R_g$ is the chain radius of gyration $R_g^2=Na^2/6$. Similarly, $s$ is rescaled by $N$, yielding
$r\to r/\ell_{0}$, $s\to s/N$, $\chi\to N \chi$,
$\omega_{j}(r)\to N\omega_{j}(r)$ and $u_{j}(r)\to N u_{j}(r)$ with $j=$A or B. With
this rescaling, we  rewrite the self-consistent equations as:
\begin {eqnarray}
\omega_{A}(r)&=&\chi\phi_{B}(r)-u_{A}(r)+\eta(r)\label{f5}\\
\omega_{B}(r)&=&\chi\phi_{A}(r)-u_{B}(r)+\eta(r)\label{f6}\\
\phi_{A}(r)&=&\frac{1}{Q_{C}}\int^{f}_{0} ds\,q_{A}(r,s)q^{\dag}_{A}(r,f-s)\label{f7}\\
\phi_{B}(r)&=&\frac{1}{Q_{C}}\int^{1-f}_{0} ds\,q_{B}(r,s)q^{\dag}_{B}(r,1-f-s)\label{f8}\\
\frac{\partial q_{j}(r,s)}{\partial s}&=&\left(\frac{R_{g}}{\ell_{0}}\right)^{2}
\nabla^{2}q_{j}\left(r,s\right)-\omega_{j}(r)q_{j}(r,s)\label{f8a}
\end {eqnarray}
where $f=N_{A}/N$,
$\Omega_1\equiv \Omega/\ell^{2}_{0}$ and $Q_{C}=(1/{\Omega_1})\int {\rm d}^2r\, q^{\dag}_{A}(r,f)$.
Note that the incompressibility condition, Eq.~(\ref{f2a}), together with Eqs.~(\ref{f5}) and (\ref{f6})
can be used to obtain the Lagrange multiplier $\eta(r)$

\begin{equation}\label{f8b}
\eta(r)=\frac{1}{2}(\omega_A+\omega_B-\chi+u_A+u_B)
\end{equation}

With the rescaled variables, we define now
a rescaled free energy:
\begin{eqnarray}\label{f11}
\frac{N a^2}{\Omega_1\ell^{2}_{0}}\frac{F }{k_{B}T}
&=&\frac{F}{n k_{B}T}=\frac{1}{\Omega_1}\int {\rm d}^{2} r
\left[\chi\phi_{A}(r)\phi_{B}(r)-\omega_{A}(r)\phi_{A}(r)-\omega_{B}(r)\phi_{B}(r)\right]
\nonumber\\
&-& \ln Q_{C}-\frac{1}{\Omega_1}\int {\rm d}^2r \left[u_{A}(r)\phi_{A}(r)+u_{B}(r)\phi_{B}(r)\right]
\nonumber\\
&  + &\frac{1}{\Omega_1}\int {\rm d}^2 r\, \eta(r)[\phi_A(r)+\phi_B(r)-1]
 \end{eqnarray}
The above self-consistent equations can be solved numerically in the
following way. First, we guess an initial set of values for the
 auxiliary fields $\omega_{j}(r)$. Then, through the diffusion
equations, Eq.~(\ref{f8a}),
we calculate the propagators, $q_j$ and $q^\dagger_j$.
Next, we calculate the monomer volume
fractions $\phi_j$ from Eqs.~(\ref{f7})-(\ref{f8})  and  the Lagrange multiplier $\eta(r)$
from Eq.~(\ref{f8a}). We can now proceed with a new set of values for $\omega_{j}(r)$
obtained through Eqs.~(\ref{f5})-(\ref{f6}), and this procedure
can be iterated until convergence is obtained by some conventional
criterion described below.

We use the semi-implicit relaxation scheme~\cite{fredrickson} to solve the two-dimensional modified diffusion equations,
Eq.~(\ref{f8a}). Our convergence criterion is based on the incompressibility condition. For perfect structures such as
parallel or perpendicular lamellae, the maximum allowed deviation between the sum of the A and B densities and unity,
$|1-\phi_A(r)-\phi_B(r)|$, is $10^{-4}$, whereas for the mixed L$_{\rm M}$  phase (see below), it is around  $10^{-2}$.
As mentioned above we rescale all lengths by the natural periodicity of the BCP, $\ell_0\simeq 4.05R_g$, and the
curvilinear coordinate, $s$, by the total number of monomers in one chain, $N$. The spatial discretization in the
$x$-direction is 0.05 (in units of $\ell_0$), while in the $z$-direction it is 0.025. The discretization of the $s$
variable is 0.02. For all the presented results, the free energy changes in the last few iteration steps are less
than $10^{-4}$ in units of $k_BT/{\rm chain}$ after the first 1,000 iterations and decreases to $10^{-6}$ after
additional 4,000 iterations. Note that since we work at a mean-field level (SCFT), it would not be of advantage to
further refine the convergence of the free energies to a higher accuracy, since we neglect anyway quadratic fluctuations
that might give larger corrections.

\section{Results} \label {sec. 4}

We present now the numerical results for symmetric di-block films ($f=1/2$) at
various patterned surfaces. The natural periodicity of the BCP,
$\ell_{0}$, is chosen for all the numerical calculations to be $50$\,nm. This value roughly corresponds to
values used in several experimental set-ups \cite{r12,r15a,r15b}. All lengths are
rescaled by $\ell_0$ as was explained in Sec.~III. Except when explicitly mentioned,
all results are obtained by using the fully disordered phase of the BCP film,
$\phi_A(r)=\phi_B(r)=0.5$, as initial condition. Then, a temperature quench
is performed from the disordered state above the ODT to
temperatures below the ODT where the lamellar phase is stable.

\subsection{Chemically striped surface}
\begin{figure}[htp]
  \begin{center}
    {\resizebox{3.0in}{!}{\includegraphics[bb=-10 0 357 310,angle=0,draft=false]{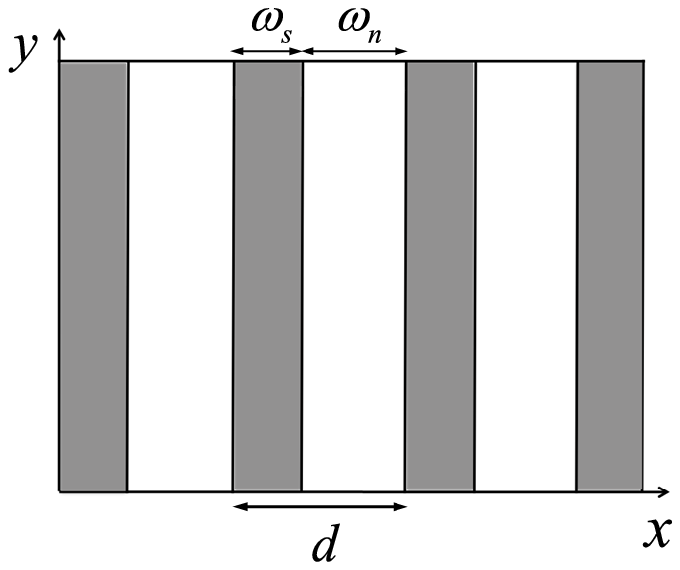}}}
    \caption{\textsf{Top view of a striped surface
 lying in the $x-y$ plane. The periodicity is $d=\omega_s+\omega_n$,
where the A-preferring stripes have a width
 of $\omega_s$ and the neutral inter-stripe regions are of  width $\omega_n$.
 \label{fig3}}}
  \end{center}
\end{figure}

The system is modeled using a SCFT scheme  for two separate set-ups that are
motivated by recent experiments \cite{r9,r11,r12,r15a,r15b}. In the first set-up, the BCP film is
spread on a {\it flat but chemically patterned} solid surface, while
the second bounding surface is the free film/air interface, which is
either neutral or has a slight preference towards one of the two BCP
components. In our calculations we take this top surface to be always neutral.
A top view of the bottom patterned surface can be seen
in Fig.~\ref{fig3}, and is composed of infinitely long stripes in the
$y$-direction of width $\omega_s\sim 100$\,nm that prefer the A
component ($\delta u>0$).
These stripes are separated by neutral inter-stripe regions of
width $\omega_n$ having the same affinity for A and B ($\Delta u=0$).
As the stripes are infinitely long in the $y$-direction, the
chemical surface pattern has a one-dimensional square-wave shape and is periodic
in the $x$-direction, $\Delta u(x+d)=\Delta u(x)$, with periodicity $d=\omega_s+\omega_n$
\begin{eqnarray}
\Delta u(x)& =& u_s ~~~{\rm for}~~~ 0< x\le \omega_s\nonumber \\
\Delta u(x) &= &u_n ~~~{\rm for}~~~ \omega_s< x\le d
\end{eqnarray}
Note that we can write formally the surface preference field
$\Delta u=u_A-u_B$  as $\Delta u(r)=
\Delta u(x)\delta(z)$, where $\delta(z)$ is the Dirac delta function.  All numerical values
of $\Delta u$ are given hereafter in terms of its rescaled units, $\Delta u \to N\Delta u$.

\begin{figure}[htp]
  \begin{center}
    {\resizebox{3.5in}{!}{\includegraphics[angle=0,scale=0.75,draft=false]{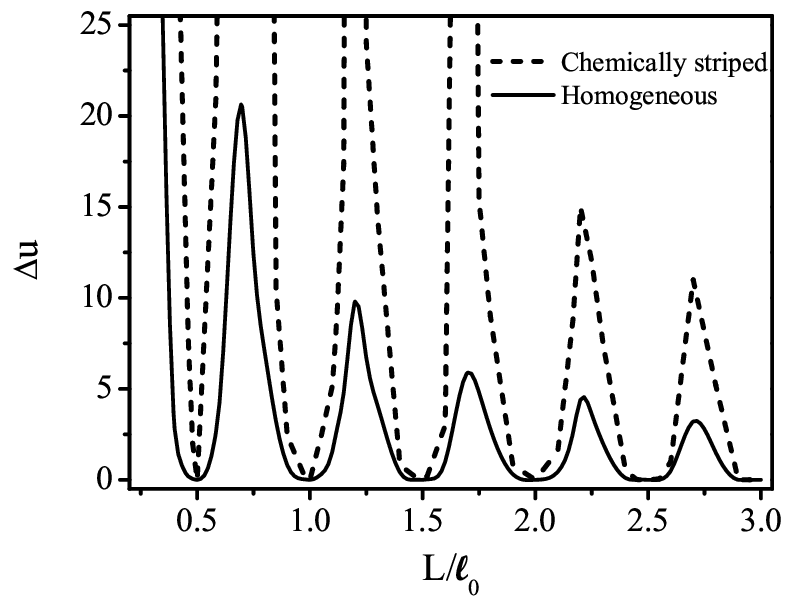}}}
  \caption{\textsf{Phase diagram in terms of the film thickness $L$ vs the surface preference
$\Delta u$ for chemically striped
surface (dashed line) and homogeneous surface (solid
line). The lines separate the parallel phase (L$_{||}$) for larger $\Delta u$ values from perpendicular one
(L$_\perp$) at smaller $\Delta u$ values, although the effective $\Delta u$ on the striped surface is
$\Delta u_{\rm eff}=\Delta u (\omega_s/d$).
The parameters used are $N \chi=20$, $\ell_0=50$\,nm and for the striped surface: $\omega_s=2\ell_0$,
$\omega_n=8\ell_0$
so that $d=10\ell_0=500$\,nm.
\label{fig4}}}
  \end{center}
\end{figure}

In the following we fix the width $\omega_s$ to be
twice the natural periodicity, yielding $\omega_s{=}2\ell_0{=}100$\,nm. The
phase diagram shown in Fig.~\ref{fig4} is calculated in terms of the film thickness,
$L$, and the bottom surface preference, $\Delta u$, for this set-up and compared with the one in
Fig.~\ref{fig2} for homogeneous surfaces. All parameters here are taken
to be the same as for the homogeneous surface, except that the bottom surface has
chemical stripes. Furthermore, we fix the value of the inter-stripe distance (where there is no
preferred adsorption), to $\omega_n=8\ell_0=400$\,nm so that the pattern periodicity is
$d=\omega_s+\omega_n=500$\,nm, or $10\ell_0$. The phase diagram is obtained by starting as an initial
guess from the perpendicular lamellar phase (${\rm L}_{\perp}$)
or the parallel one (${\rm L}_{||}$). After convergence, their free energies is compared.
From the figure it is evident that the ${\rm L}_{\perp}$ phase has a larger stability range
for the chemically striped surface as compared with the
homogeneous surface, although the effective value of $\Delta u$ on the entire patterned
surface is smaller as its value should be averaged over both the striped and inter-stripe regions:
$\Delta u_{\rm eff}=\Delta u(\omega_s/d)$. Note that the stability of the L$_\perp$ phase is in particular enhanced
for special values of $L$:
$\frac{3}{4}\ell_0, \frac{5}{4}\ell_0\dots$.

\begin{figure}[htp]
  \begin{center}
    {{\includegraphics[bb=35 0 220 330,angle=0,scale=1.5,draft=false]{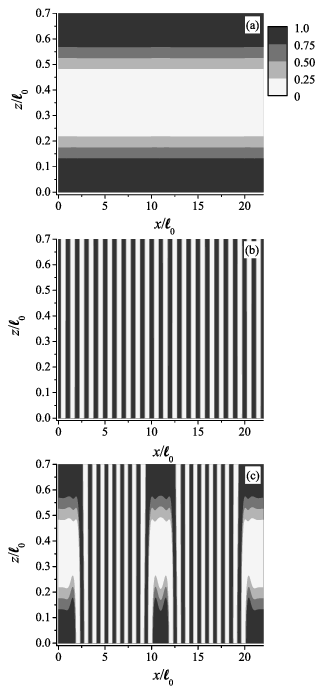}}}
\caption{\textsf{BCP lamellar structures obtained from numerical solutions of SCFT equations
for three different initial conditions: (a)~parallel lamellar (${\rm L}_{||}$);
(b)~perpendicular lamellar (${\rm L}_{\perp}$); and, (c)~fully disordered state developing into
a mixed morphology (${\rm L_M}$). The film thickness is $L=0.7\ell_0$,
the top surface is taken as neutral,
$\Delta u=0$, while the bottom one has a striped pattern as in Fig.~\ref{fig3} with $\Delta u=1$.  The inter-stripe
widths are set to be $\omega_s=2\ell_0$ and $\omega_n=8\ell_0$, yielding
$d=\omega_s+\omega_n=10\ell_0=500$\,nm. The other parameters are $N \chi=20$ and $\ell_0=50$\,nm. The color code
corresponds to the four intervals of local monomer density $0\le\phi_A(r)\le 1$, as is depicted in part (a).
\label{fig5}}}
  \end{center}
\end{figure}

\begin{figure}[htp]
  \begin{center}
    {\resizebox{3.2in}{!}{\includegraphics[bb=0 0 342 293,angle=0,draft=false]{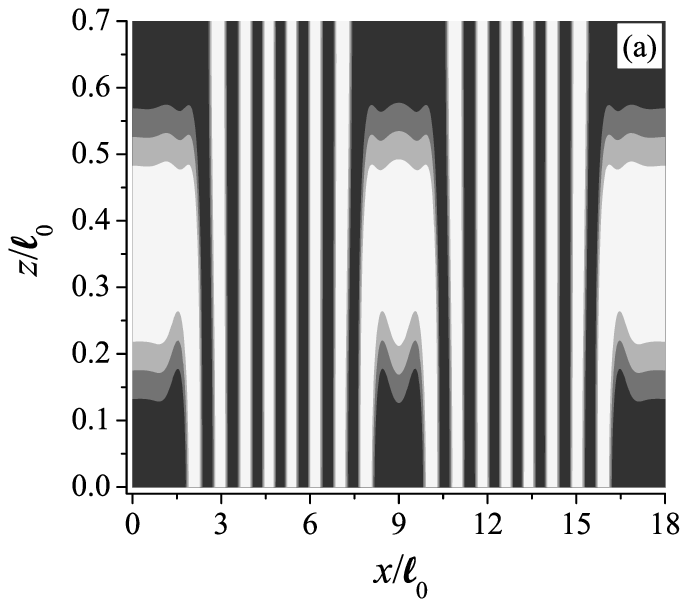}}}
    {\resizebox{3.2in}{!}{\includegraphics[bb=0 0 342 293,angle=0,draft=false]{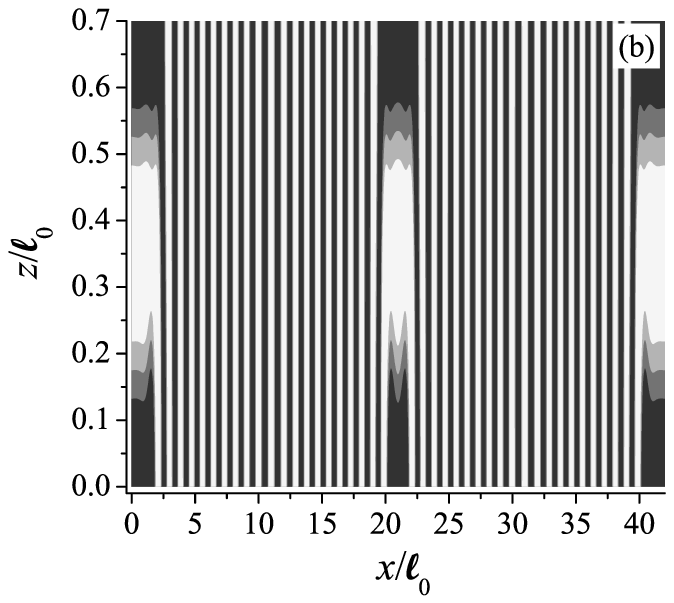}}}
  \caption{\textsf{Calculated BCP lamellar structures
  for patterned surfaces of increasing inter-stripe
  distance while $\omega_s=2\ell_0$ remains fixed (and, hence, increasing $d$). In (a)
  $\omega_n=6\ell_0$ and $d=8\ell_0=400$\,nm; in (b) $\omega_n=18\ell_0$ and $d=20\ell_0=1$\,$\mu$m.
  All lengths are scaled with the lamellar periodicity $\ell_0=50$\,nm. The other parameters are:
  $L=0.7\ell_0$, $\Delta u=1$, and $N \chi=20$.
  The  initial condition is chosen as the fully disordered state, $\phi_A(r)=0.5$ and all other parameters and color code
  are the same as in Fig.~\ref{fig5}. The system exhibits mixed ${\rm L_M}$ morphologies
  with L$_{||}$ regions just on top of the
  surface stripes and perfect L$_{\perp}$ domains in between the stripes.}
 \label{fig6}}
  \end{center}
\end{figure}

Due to the existence of many metastable states in BCP melts, the numerical procedure of free energy minimization
is sensitive to the initial conditions.
Instead of converging always to the true equilibrium structure at any point of the phase diagram,
different metastable structures can be obtained.
We show some results to illustrate this scenario in Fig.~\ref{fig5}.

For $L=0.7\ell_0$, $d=10\ell_0$, $\Delta u=1$ and $N \chi=20$ (a typical
set of parameters that is located inside the ${\rm L}_{\perp}$ stable region),
we start with parallel lamellae, perpendicular lamellae and
the fully disordered state as three different initial conditions
and perform a temperature quench  to a
temperature below the ODT. The ${\rm L}_{||}$ and ${\rm L}_{\perp}$ phases in Fig.~\ref{fig5}a and
\ref{fig5}b, respectively, result from quenching from
${\rm L}_{||}$ and ${\rm L}_{\perp}$ initial conditions. Hence, the system retains its orientation
after the temperature quench. On the other hand, for the fully disordered initial condition, we
obtain a mixed structure containing domains of the ${\rm L}_{||}$ and ${\rm L}_{\perp}$ phases.
This structure is shown in Fig.~\ref{fig5}c and is coined as ${\rm L_M}$.

As explained in Sec.~III, the maximal deviation of the incompressibility condition, $|1-\phi_A(i,j)-\phi_B(i,j)|$, serves
as our accuracy criterion. It is $1.10\times 10^{-6}$ for the parallel lamellae as initial condition (Fig.~\ref{fig5}a); $2.57\times 10^{-5}$ for
the perpendicular lamellae as initial condition (Fig.~\ref{fig5}b); and,  $1.07\times 10^{-2}$
for disordered state as initial condition (Fig.~\ref{fig5}c).
For the L$_{||}$ and L$_\perp$ it is quite small, yielding a value of about $10^{-5}$. However,
it is not as good in the mixed L$_{\rm M}$ structure ($10^{-2}$), because of the existence of internal boundaries
between parallel and perpendicular domains.
To answer the question of metastability we calculate the free-energies per chain and obtain  $f_{||}=4.272 > f_{\rm M}=4.122> f_\perp=4.061$,
corresponding to the ${\rm L}_{||}$, ${\rm L_M}$ and ${\rm L}_{\perp}$ phases, respectively.
Clearly, the most stable structure is the perpendicular one, ${\rm L}_{\perp}$,
and is consistent with our phase diagram in
Fig.~\ref{fig4}. Note that the free energy differences between the various states
are very small, on the order of 2-5\%,
manifesting the tendency of the system to get trapped into metastable states.

Our findings have also experimental implications because in experiments the film
structure depends strongly on its history and sample preparation \cite{r15b,r29a,r29b}.
The claim is that once the system is prepared in its ${\rm L}_{\perp}$ it will stay there.
But if the film is prepared above
the ODT, in its fully disordered state, the film can get stuck in a metastable mixed lamellar structure,
${\rm L_M}$.  Although in experiments it is not always possible to heat the system above its
ODT because of polymer break-down and oxidation, in many cases, higher temperatures are used
to anneal the film and allow it to reach its final state via faster dynamics.

Another interesting feature is presented in Fig.~\ref{fig5}. Perfect
perpendicular lamellar structures between neighboring stripes are
visible. Furthermore, we can obtain such perfect ${\rm L}_{\perp}$
structures for a wide range of small and large periodicities, ranging from
$d=400$\,nm in Fig.~\ref{fig6}(a) to $d=
1\,\mu{\mathrm m}=20\ell_0$ in Fig.~\ref{fig6}(b).
However, we find that it is difficult to get rid of the parallel
lamellar regions induced by the striped pattern, even when we
further reduce the BCP film thickness $L$ to values much less than
$\ell_{0}$ and decrease the values of $\Delta u$.
Furthermore, a preliminary study \cite{tobepublished} indicates that slow temperature annealing
from the disorder state (above ODT) to the ordered lamellar state (below ODT) does not seem to prevent
the formation of the mixed ${\rm L_M}$ phase.

\subsection{Periodic grooved surfaces}
\begin{figure}[htp]
  \begin{center}
    {\resizebox{4.0in}{!}{\includegraphics[bb=20 40 331 248,angle=0,scale=0.7,draft=false]{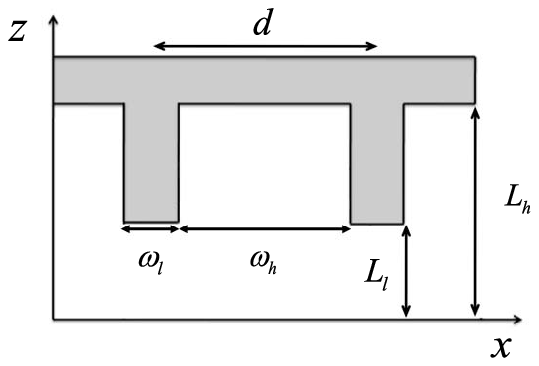}}}
    \caption{\textsf{A cut (side view) through the top grooved surface (the mold).
    The periodicity in the $x$-direction is $d=\omega_l+\omega_h$, with $\omega_l$ and $\omega_h$ being the finger and inter-finger width,
    respectively. $L_l$ is the distance of closest approach to the bottom surface at $z=0$, and  $L_h$ is the largest film
thickness. The initial film thickness is equal to the average film thickness in the mold, $L=(L_l\omega_l+L_h\omega_h)/d$.
\label{fig7}}}
  \end{center}
\end{figure}

In order to overcome the problem of getting trapped in ${\rm L_M}$ mixed states and inspired by
recent NanoImprint lithography (NIL) experiments~\cite{r15b,r29,r29a,r29b}, we explored yet another type of surfaces.
The set-up can be seen in Fig.~\ref{fig7}, where the BCP film is confined between two solid surfaces.
The bottom surface at $z=0$ is flat, while
the top one has a periodic arrangement of grooves (along the $x$-direction) made of a series
of down-pointing `fingers' of thickness $\omega_l$, separated by
inter-grooves regions (`plateaus') of thickness $\omega_h$. The
periodic height profile $h(x)=h(x+d)$ has the form:
\begin{eqnarray}
h(x)&=&L_l~~~ {\rm for}~~~ 0< x \le \omega_l \nonumber\\
h(x)&=&L_h~~~ {\rm for}~~~ \omega_l< x \le d
\end{eqnarray}
where the height is measured from the $z=0$ surface. Formally, $\Delta u(r)=u_A-u_B$ used in the solution of Eqs.~(\ref{f5})-(\ref{f6})
is given by $\Delta u(r)=\Delta u(x)\delta(z-h(x))$.

The figure shows  the surface height profile
in the $x-z$ plane, for profiles that are translationally
invariant in the $y$-direction.  The periodicity
in the $x$-direction is  $d=\omega_l+\omega_h$ and
the finger width is chosen to be $\omega_l=5\ell_0=250$\,nm.
The top surface (mold) is put in
direct contact with a BCP film spread on a neutral and flat bottom
surface (at $z=0$). The distance of closest approach between the two
surfaces is $L_l$, while the maximal height difference between them
is $L_h$. This means
that the finger height of the mold is $L_h-L_l$.
Assuming film incompressibility, we get a relation between the
thickness $L$ of the original BCP film and the two
height parameters, $L_l$ and $L_h$: $L=(L_l \omega_l +
L_h \omega_h)/d$. In experiments, the average thickness $L$ is fixed,
while in the numerical study, we control directly
$L_l$ and $L_h$.

By varying the values of the parameters $d$, $L_h$ and $L_l$ of
the mold, and the strength of surface interactions $\Delta u$, we can get a
sequence of BCP patterns. Furthermore, we obtain perfect
perpendicular lamellar structures extending throughout the film
thickness for some special patterned surfaces.

\begin{figure}[htp]
  \begin{center}
    {\resizebox{3.5in}{!}{\includegraphics[angle=0,scale=0.75,draft=false]{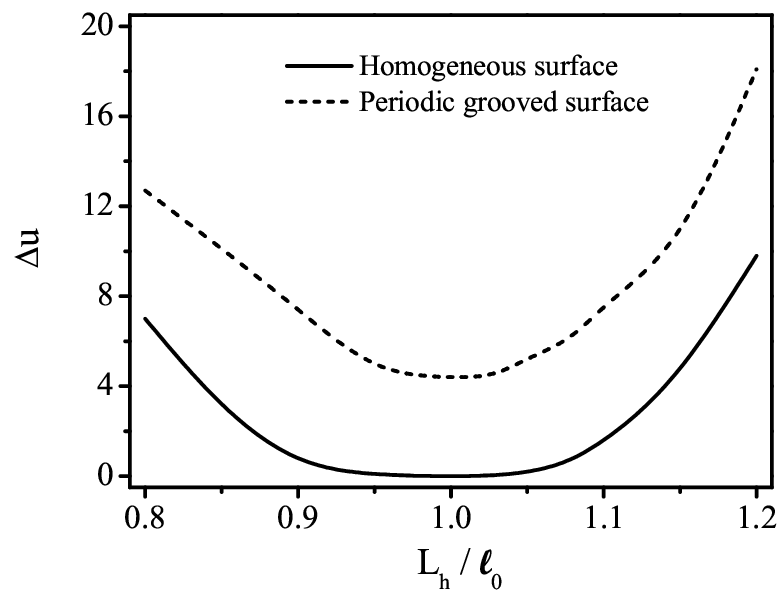}}}
  \caption{\textsf{Phase diagram in terms of the maximal film thickness
  $L_h$ vs the surface field difference of the two blocks, $\Delta
  u$, for a periodic grooved surface (dashed line) and a homogeneous
  surface (solid line). In the latter case, the film thickness $L$ is equated with $L_h$.
   Other parameters are:
  $\ell_0=50$\,nm,
  $L_l=0.3\ell_{0}$, $d=15\ell_{0}=750$\,nm, $\omega_l=5\ell_0=250$\,nm, $\omega_h=10\ell_0=500$\,nm and $N \chi=20$.
 \label{fig8}}}
  \end{center}
\end{figure}

We calculate the phase diagram in terms of the maximal film thickness,
$L_h$, vs the surface preference, $\Delta u=u_A-u_B$.  The interaction strength on all exposed surfaces
of the
upper grooved mold have the same value of $\Delta u$. In addition, we set
$L_l=0.3\ell_{0}$, $d=10\ell_{0}$ and $\omega_l=5\ell_0$. The result is shown in Fig.~\ref{fig8}, from which
we can infer that this set-up greatly affects the phase diagram as compared with Fig.~\ref{fig2}
for a uniform $\Delta u$ surface. The transition line from  L$_{||}$
to L$_\perp$ is shifted upwards so that its minimum is obtained for $L_h=\ell_0$ where
$\Delta u=4.4$. This is similar but more pronounced than the behavior seen in Fig.~\ref{fig4} for the
chemical striped surface around the $L/\ell_0=1.0$ region.

\begin{figure}[htp]
  \begin{center}
    {\resizebox{3.5in}{!}{\includegraphics[bb=0 0 342 293,angle=0,draft=false]{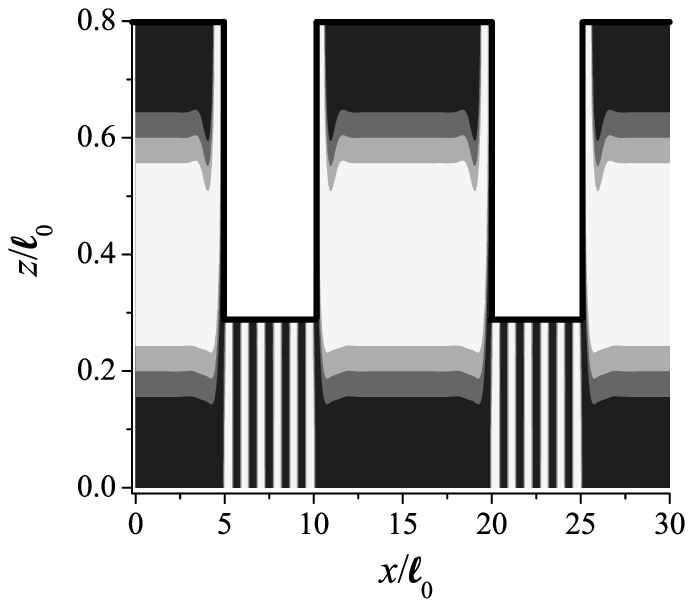}}}
  \caption{\textsf{BCP density distribution for $d=15\ell_0=750$\,nm and
  $N \chi=20$ starting from a fully disordered initial condition.
The bottom surface is neutral and the top surface has a square wave
height profile as in Fig.~\ref{fig8} where $\Delta u=0.1$.
Other parameters are $L_h=0.8\ell_{0}$ and $L_l=0.3\ell_{0}$, $\omega_h=10\ell_0$, $\omega_l=5\ell_0$,
yielding $L/\ell_0=1.9/3\simeq 0.64$.
 \label{fig9}}}
  \end{center}
\end{figure}

However, when we start from a fully disordered
state as initial condition inside the stable ${\rm L}_{\perp}$ region of Fig.~\ref{fig8}
({\em e.g.,} $L_h=0.8\ell_{0}$ and $\Delta u=0.1$), we do not
get the fully perpendicular lamellae ${\rm L}_{\perp}$ but rather a mixture of parallel and
perpendicular lamellar regions (the ${\rm L_M}$ structure) as shown in Fig.~\ref{fig9}.

\begin{figure}[htp]
  \begin{center}
    {\resizebox{3.in}{!}{\includegraphics[bb=0 0 342 293,angle=0,draft=false]{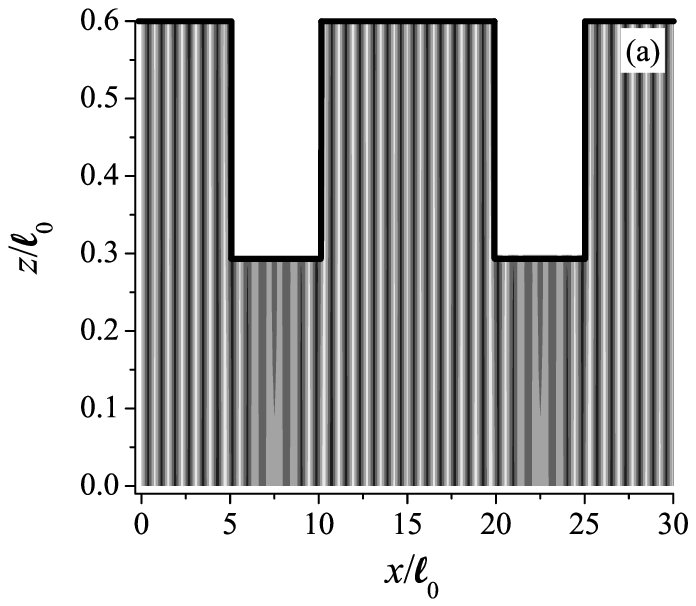}}}
    {\resizebox{3.in}{!}{\includegraphics[bb=0 0 342 293,angle=0,draft=false]{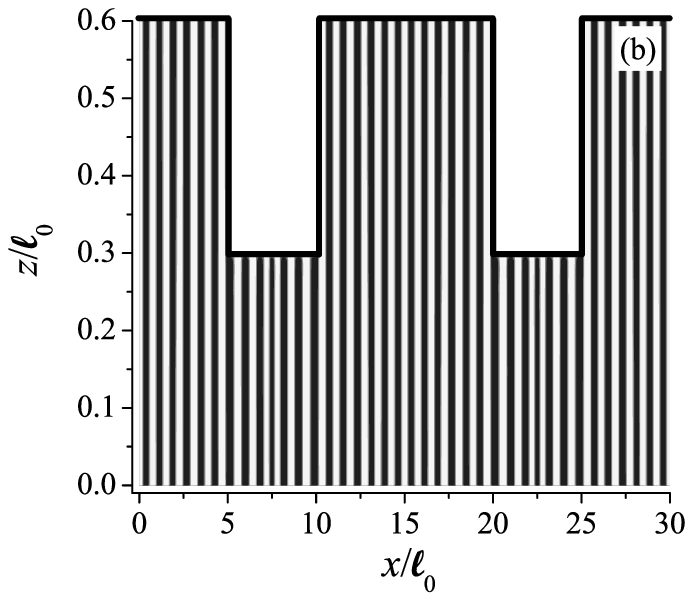}}}
\caption{\textsf{BCP density distribution for $d=15\ell_0=750$\,nm and $N \chi=11.5$ when the
initial condition is the fully disordered state. The system is first annealed to $N\chi=11.5$ in (a)
and then to $N \chi=20$ in (b).
The bottom surface ($z=0$) is neutral and the top surface has a square
grooved structure with $\Delta u=0.1$. Other parameters are
$L_h=0.6\ell_{0}$, $L_l=0.3\ell_{0}$, $\omega_h=10\ell_0$ and $\omega_l=5\ell_0$, yielding $L=0.5\ell_0$.
 \label{fig10}}}
  \end{center}
\end{figure}

\begin{figure}[htp]
  \begin{center}
    {\resizebox{3.in}{!}{\includegraphics[bb=0 0 342 293,angle=0,draft=false]{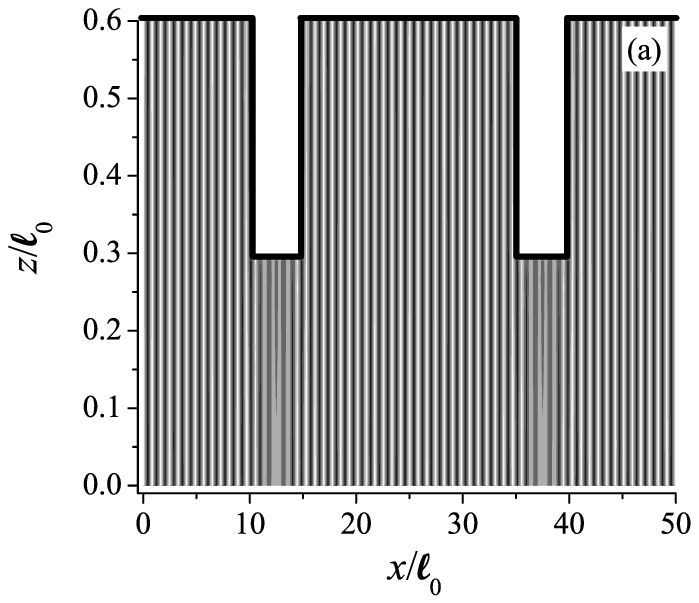}}}
    {\resizebox{3.in}{!}{\includegraphics[bb=0 0 342 293,angle=0,draft=false]{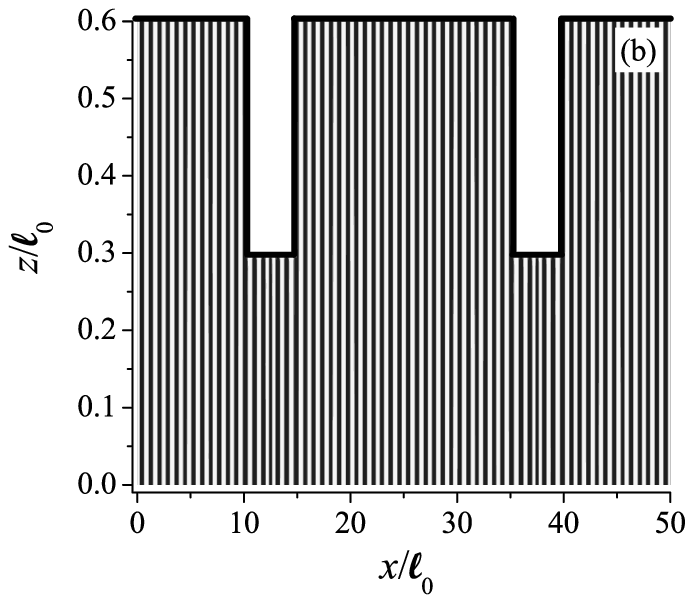}}}
  \caption{\textsf{BCP density distribution for $d=25\ell_0=1.25$\,$\mu$m using two-step annealing procedure.
  First to $N \chi=11.5$ in (a)
  and then  to $N \chi=20$ in (b). The initial condition is the fully disordered state. Other parameters are:
  $L_h=0.6\ell_0$, $\omega_l=5\ell_0$, $\omega_h=20\ell_0$ and $L_l=0.3\ell_0$, yielding $L=0.54\ell_0$.
 \label{fig11}}}
  \end{center}
\end{figure}

We find two ways to improve on the perpendicular orientation by
changing the mold geometry and surface characteristics. First, we
decrease the film thickness by decreasing $L_h$ to $0.6\ell_{0}$. In
this case, we do a gradual temperature quench, starting from the
disordered state above the ODT and quenching to temperatures just
below the ODT, $N \chi=11.5$, and only then proceed with a deep quench
to $N \chi=20$. This two-step  procedure is shown in
Fig.~\ref{fig10}(a) and (b). Perfect perpendicular lamellar structures emerge.
Moreover, using this two-step procedure we can even obtain a
perfect perpendicular lamellar structures with much wider $\omega_n$ yielding
$d=1.25$\,$\mu$m (or equivalently $d/\ell_0=25$),
as is shown in Fig.~\ref{fig11}.

\begin{figure}[htp]
  \begin{center}
    {\resizebox{3.in}{!}{\includegraphics[bb=0 0 342 293,angle=0,draft=false]{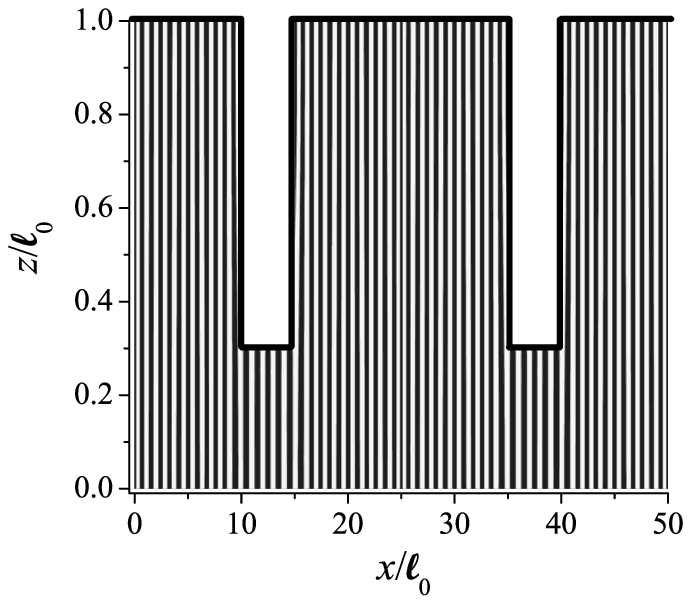}}}
  \caption{\textsf{BCP density distribution for $d=25\ell_0=1.25$\,$\mu$m.
The bottom surface at $z=0$ is neutral, $\Delta u=0$. The top surface is a square grooved
with $\Delta u=0.1$ on the sides and tips of the grooves  and neutral ($\Delta u=0$) on the top plateau parts (see text)
and Eq.~(\ref{ff18}).
Other parameters are $N \chi=20$, $L_h=\ell_{0}$, $L_l=0.3\ell_{0}$, $\omega_h=20\ell_0$ and
$\omega_l=5\ell_0$, yielding $L\simeq 0.86\ell_0$.
 \label{fig12}}}
  \end{center}
\end{figure}

A second variation is to  construct the grooves from two separate
materials with different A/B preference. The protruding `finger' parts
are assumed to have a small A preference ($\Delta u=u_1>0$)
both on their vertical and horizontal parts,
while the high plateau parts are taken as neutral ($\Delta u=0$).
\begin{eqnarray}
\Delta u(x)&=& u_1~~~{\rm for}~~~0\le x\le \omega_s\nonumber\\
\Delta u(x)&=& 0~~~~{\rm for}~~~\omega_s< x< d\label{ff18}
\end{eqnarray}
With this special surface geometry and interactions, we obtain perfect
${\rm L}_{\perp}$ structures for wide range of film
thicknesses. An example for such a set-up with $d=25\ell_0$ periodicity is shown in
Fig.~\ref{fig12}.

\section{Discussion and Conclusions}

In this paper we addressed several surface patterns as
inspired from recent experiments in relation with ordering
and orientation of lamellar phases of block copolymer (BCP) films.
In the first
set-up, we model a BCP film  confined between a chemical
striped solid surface and the free film/air
surface. In a second set-up, the film is considered to occupy the gap
between
two solid surfaces; a  flat one and a hard mold with specific
square-shape grooves as is inspired from recent NanoImprint lithography (NIL) experiments.

The main question both  experiments and  modeling should attempt answering is
how to induce a perfect perpendicular order
in BCP films? In particular, how this can be achieved
using patterned surfaces with structural features (stripes and grooves)
that have a periodicity $d$ much larger than the
lamellar periodicity $\ell_0$. Having such sparse surface features
 will reduce substantially the cost of large-scale production of  surface templates and BCP films
and is essential for applications, {\it e.g.,} in microelectronic and nano-lithography processes.

Using the first set-up of the chemically stripes on an otherwise flat and neutral surface, we are able
to show that the perpendicular phase ${\rm L_\perp}$ has a larger stability region in parameter space
described by the film thickness $L$ and  surface preference ($\Delta u$), as compared with the
homogeneous surface. Note that this is the case even for inter-stripe distances $\omega_n$
that are an order of magnitude larger than the stripe thickness, $\omega_s$. This is in spite the fact
that the effective (averaged) $\Delta u$ for the striped surface is smaller than the corresponding
$\Delta u$ on the homogeneous surface, $\Delta u_{\rm eff}=\Delta u(\omega_s/d)<\Delta u$.
We equally find that the system is very sensitive to initial conditions.
Starting from a fully disordered state, above the order-disorder temperature (ODT)
and annealing the temperature into the lamellar region, will
mainly produce a mixed morphology ${\rm L_M}$ as can be seen in Fig.~\ref{fig5}(c) and Fig.~\ref{fig6}.
Although the stripes nucleate growth of BCP layers on top of them (namely, domains with  parallel orientation,
${\rm L_{||}}$),  perfectly oriented perpendicular domains, ${\rm L_\perp}$, are induced on top of the neutral inter-stripe region.

In our model the L$_{\rm M}$  mixed morphology is a result of the large number of metastable states (local minima)
which the system possesses. Although the true equilibrium is the ${\rm L_\perp}$ phase, it is hard to find it
numerically unless one starts with  the proper initial conditions. This drawback should also be expected in experiments,
where during sample preparation, the film undergoes many external stresses and defects are abundant. It will be of interest to verify in experiment
our findings by doing a slow temperature
annealing of BCP films from their disordered liquid state (above ODT) into the lamellar region (below the ODT).
Such a slow temperature annealing has the potential to produce highly oriented BCP films. Although not in all systems
it is possible to reach temperatures above the ODT without damaging the BCP chains, we equally note that in many cases
 annealing at high enough temperatures has the advantage that the system can reach it final state with faster kinetics.

In the second set-up, we modeled a hard mold which is pressed onto a BCP lamellar film. We show that this NanoImprint
lithography (NIL)  process greatly enhances perpendicular order in lamellar phases. Perfect ${\rm L_\perp}$ can be
seen for film thicknesses below $\ell_0$ even when the groove width $\omega_h$ (filled with the BCP
film) is five times (or even larger) than the solid `finger' ($\omega_l$) sections. Here the slow
annealing from above ODT to below the ODT is very successful, demonstrating that
this set-up is more suitable  for lamellar orientation purposes than the chemical stripe set-ups discussed above.

In Fig.~\ref{fig12} we proposed a mold with even superior orientation qualities. For this mold the surface preference of the downward
protrusion sections (the `fingers') is larger than that of the top section of the groove (plateau-like). As the latter
 preference interferes with the  ${\rm L_{||}}$ ordering, reducing this surface preference will enhance ${\rm L_\perp}$
 ordering, especially in the desired case of thin fingers and wide plateaus, where $\omega_h\gg \omega_l$.

\begin{figure}[htp]
  \begin{center}
    {\resizebox{3.in}{!}{\includegraphics[width=10cm,draft=false]{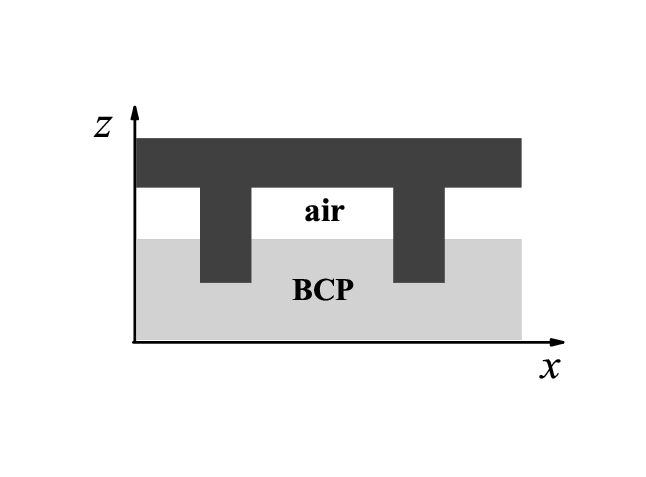}}}
  \caption{\textsf{Schematic drawing of the NanoImprint lithography (NIL) set-up where the mold only partially
  is filled with the BCP film. Effectively, this means that the film sections in contact with the
  side boundaries of mold feel a different surface field than
   the top horizontal facets, which are exposed to the air.
 \label{fig13}}}
  \end{center}
\end{figure}

In experiments, it is harder to produce a mold with such specific surface characteristics as seen in Fig.~\ref{fig12}.
One way would be to form it from two separate materials or to use a selective coating during mold preparation. However,
 creating such a mold can be a costly and delicate process that will be hard to mass reproduce. Yet another possibility
 is to have an {\it effective} chemically heterogeneous mold shown in Fig.~\ref{fig13}.
Suppose that the groove height is only partially filled with the BCP melt, creating pockets of air on the top
of each groove~\cite{r29}.
The film/air interface within each groove can be thought of as another interface with almost neutral
preference the two blocks. This situation amounts to taking different values of $\Delta u$ on the finger-section
and plateau-section of the mold [see Eq.~(\ref{ff18})].
While $\Delta u=0$ on the top section (plateau) of the groove, it is non-zero on the mold `finger'
sections. This is exactly the situation explored in our calculations and shown in Fig.~\ref{fig12}, and it may be
worthwhile to further explore this partial filled mold in future experiments.

Our theoretical modeling relies on numerical solutions of self-consistent field theory (SCFT) equations. We minimize the
corresponding free energies and converge to film morphology whose free energy is an extremum using an iterative procedure.
We find that the numerical procedure is sensitive to what is used
as initial conditions for the BCP structure. The
convergence can be towards local (metastable) states
and not always towards the true equilibrium.  This is an unavoidable feature of the numerical procedure.
It is not an artifact but rather reflects the true physical situation as seen in experiment. The BCP film has many
metastable states separated by energy barriers and it is hard to reach the true thermodynamical equilibrium state.
Slow annealing from above the ODT  or from high temperatures is
one way to overcome this difficulty, at least in a partial way.

It will be of great interest to further proceed and extend our
two-dimensional calculations to full three-dimensional ones. This
will require much longer computation times but will allow us  to
distinguish between perfectly oriented perpendicular lamellae and
those that stand up but which also wander around in the $x-y$ plane. For applications it is important to
have perfectly oriented ${\rm L_\perp}$ phases in the $z$-direction,
that are well aligned  in the lateral (in-plane) directions.

Although our present study is not exhaustive, it shows
many possibilities of explaining some of the experimental findings
and even points towards interesting directions for future
experiments.

\section*{Acknowledgement}
Our theoretical work was motivated by numerous discussions with
P. Guenoun and J. Daillant. We would like to thank them for sharing with us their
unpublished experimental results and for valuable comments and suggestions.
Two of us (DA, XM) would like to thank the RTRA agency (POMICO project No.
2008-027T) for a travel grant, while HO acknowledges a fellowship from the Mortimer
and Raymond Sackler Institute of Advanced Studies at Tel Aviv university.
This work was partially
supported by the U.S.-Israel Binational Science Foundation under Grant No.
2006/055 and the Israel Science Foundation under Grant No. 231/08.




%
%
%




\begin{thebibliography}{99}
\bibitem{ft1}
For off-symmetrical fraction,
$f\ne 0.5$, the equilibrium state is either asymmetric lamellae or phases of completely different
symmetry such as hexagonal phases of
cylinders or others (see Ref.~\cite{HamleyBook}).
These latter phases will not be considered in this paper.
\end{thebibliography}
\end{document}